\documentclass[aps,IOP,onecolumn,superscriptaddress,groupedaddress]{revtex4-1}  
\usepackage{graphicx,amssymb}
\usepackage{color,ulem}
\usepackage{bm}   

\newcommand{\eqref}[1]{(\ref{#1})}

\begin{document}
\title{Coupled qubits as a quantum heat switch}

\author{B. Karimi}
\affiliation{Low Temperature Laboratory, Department of Applied Physics, Aalto University School of Science, P.O. Box 13500, 00076 Aalto, Finland}
\author{J. P. Pekola}
\affiliation{Low Temperature Laboratory, Department of Applied Physics, Aalto University School of Science, P.O. Box 13500, 00076 Aalto, Finland}
\author{M. Campisi}
\affiliation{NEST, Scuola Normale Superiore \& Istituto Nanoscienze-CNR, I-56126 Pisa, Italy}
\author{R. Fazio}
\affiliation{ICTP, Strada Costiera 11, 34151 Trieste, Italy}
\affiliation{NEST, Scuola Normale Superiore \& Istituto Nanoscienze-CNR, I-56126 Pisa, Italy}

\date{\today}

\begin{abstract}
We present a quantum heat switch based on coupled superconducting qubits, connected to two $LC$ resonators that are terminated by resistors providing two heat baths. To describe the system we use a standard second order master equation with respect to coupling to the baths. We find that this system can act as an efficient heat switch controlled by the applied magnetic flux. The flux influences the energy level separations of the system, and under some conditions, the finite coupling of the qubits enhances the transmitted power between the two baths, by an order of magnitude under realistic conditions. At the same time, the bandwidth at maximum power of the switch formed of the coupled qubits is narrowed.
\end{abstract}

%\pacs{}

\maketitle

\section{Introduction}
Quantum information processing based on superconducting qubits has made considerable progress in recent years (see for example the reviews \cite {clarke, devoret}). Several platforms have been realised and the first quantum protocols have been already implemented experimentally \cite {devoret}. In this respect superconducting nano-circuits are by now considered among the most promising implementations of solid-state quantum processors.

Over the years the judicious choice of new designs and the improvement in materials and fabrication has increased the coherence properties of superconducting nano circuits by several orders of magnitude. However, beyond standard relaxation and decoherence studies, the dissipation and heat transport properties of these systems are only little investigated and poorly understood, albeit important from the practical point of view. 
The newly born field of quantum thermodynamics \cite {goold,anders} has highlighted that quantum coherence and quantum correlations can play prominent role in the design of efficient thermal processes. Several theoretical proposals have already appeared discussing heat engines based on the dynamics of nano-circuits and qubits \cite{linden, venturelli, koslov, correa, campo, niedenzu}, some with a direct implementation with superconductors \cite{hofer,marchegiani,quan,campisi1}.
On a related topic, quantum limited heat transport by phonons, photons, and electrons has been investigated in solid state circuits both theoretically \cite{pendry,schmidt,rego,ojanen,pascal} and experimentally \cite {schwab,jukkanature,timofeev,partanen,jezouin}. This large body of activity quests for the research on new designs with enhanced functionalities relying on the quantum properties of the constituent circuits. 

Here, we present a new device of this family. Our focus is to introduce the coupling between a pair of qubits as elements to control heat transport in quantum regime. We present a quantum heat switch connected to two reservoirs. Using a second order quantum master equation, we quantitatively obtain the expression for power mediated by this two-qubit system. As a result of the coherent coupling, when  power is enhanced by up to one order of magnitude the bandwidth at maximum power is boosted as compared to the case in which the two qubits are independent.
\begin{figure}[th]
\centering
\includegraphics [width=0.5\columnwidth] {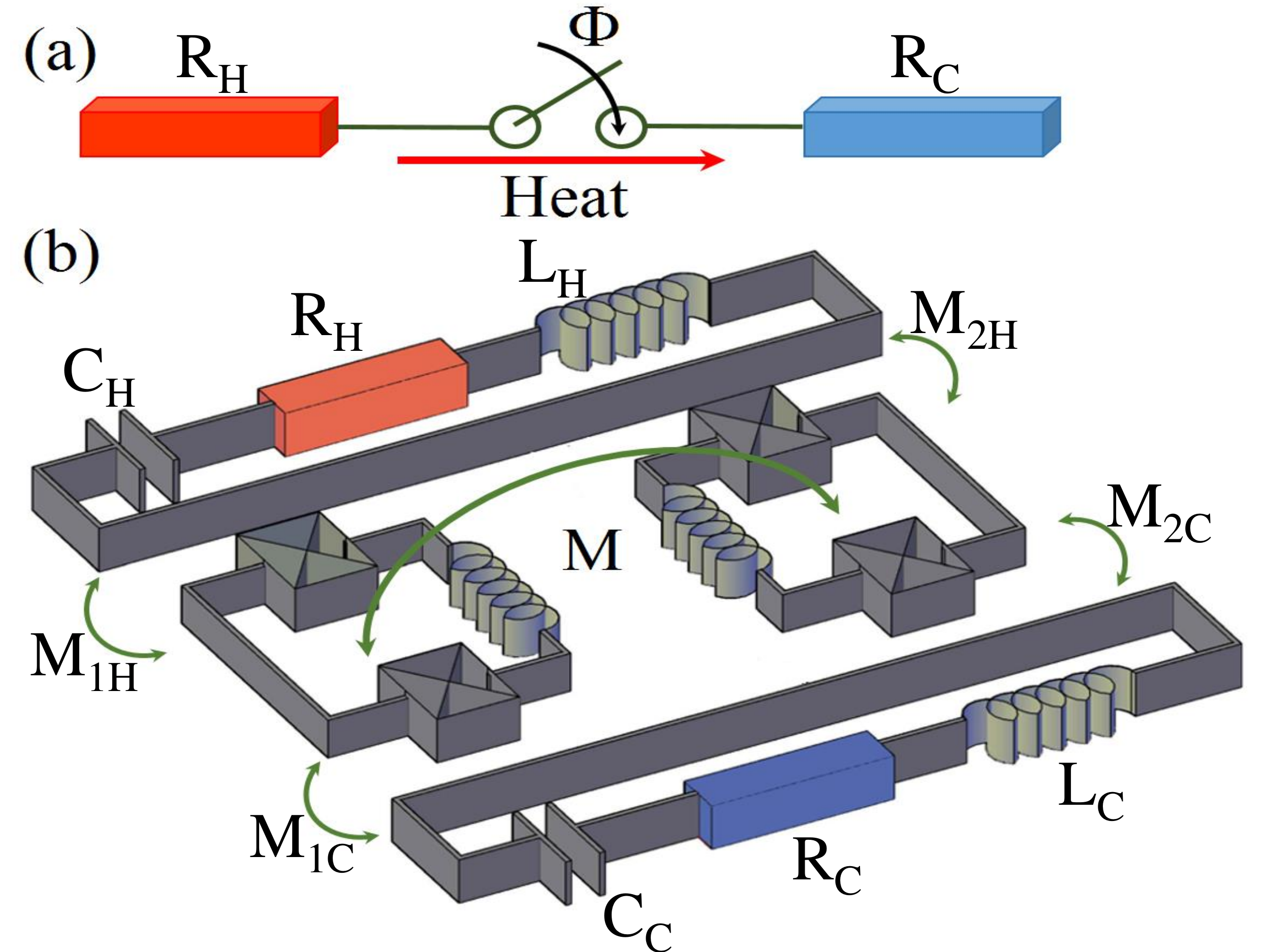}
\caption{(a) A heat switch between two baths $R_{\rm H}$ and $R_{\rm C}$ operated by control parameter $\Phi$. (b) A quantum heat switch discussed in this work.
\label{fig1}}
\end{figure}  

\section{Description of the system}\label{sec2}
The principle of a basic heat switch is shown in Fig.~\ref{fig1}a. Here, when the switch is "on", heat flows from hot bath to the cold one. In quantum circuits a heat switch can be realized by coupling the two baths by an intermediate element with tunable parameters. For instance this element can act as an energy filter thanks to its level structure. A basic element in this respect is either a classical or quantum $LC$ circuit admitting heat only at or around its resonance frequency \cite{ojanen, Jukka}. In steady state with unequal temperatures of the two baths (cold $(\rm C)$ and hot $(\rm H)$) a qubit can act as such a switch \cite{Jukka}. We show that with our suggested design (Fig.~\ref{fig1}b) composed of two coupled qubits, this operation can be boosted. In Fig.~\ref{fig1}b, the superconducting qubits in the middle are mutually coupled $(M)$ and each consists of a loop interrupted by Josephson junctions. They are further coupled to two resonators with baths ${\rm B}={\rm C},{\rm H}$ via mutual inductances $(M_{j,{\rm B}}$ for $j=1,2)$ and each resonator is a series $RLC$ circuit. Resistors $R_{\rm C}$ and $R_{\rm H}$, with temperatures $T_{\rm H}$ and $T_{\rm C}$ act as cold and hot reservoirs, respectively.
%\begin{figure}[t]
%\centering
%\includegraphics [width=\columnwidth] {circuit.png}
%\caption{a) Scheme of the quantum refrigerator presented. b) Thermodynamic Otto cycle of the refrigerator. c) configuration of the two level energies of the qubit.
%\label{fig1}
%\end{figure}
We write the total Hamiltonian of the system with its baths shown in Fig. \ref{fig1}b as
\begin{eqnarray} \label{h1}
H =H_{\rm Q1}+H_{\rm Q2}+H_{12}+ H_{\rm R_H}+ H_{\rm R_C}+H_{\rm c1,C}+H_{\rm c1,H}+H_{\rm c2,C}+H_{\rm c2,H},
\end{eqnarray}
where $H_{\rm Q1},H_{\rm Q2}$ are the Hamiltonians of the two qubits, $H_{12}$ is the coupling between them, $H_{\rm R_H},H_{\rm R_C}$ are the Hamiltonians of the hot and cold reservoirs, and $H_{\rm c1,C}, H_{\rm c1,H}$ and $H_{\rm c2,C}, H_{\rm c2,H}$ the couplings of qubits 1 and 2 to the two reservoirs. The various components of the Hamiltonian of each qubit can be written as
\begin{equation} \label{h2}
H_{{\rm Q}j} = -E_j (\Delta_j \sigma_{x,j} + q\sigma_{z,j})
\end{equation}
for $j=1,2$ with $E_j$ the overall energy scale of each qubit, $2\Delta_j$ the dimensionless energy splitting at $q = 0$, and $\sigma_{x,j},\sigma_{z,j}$ the Pauli matrices. The flux $\Phi$ is applied to the qubits using control parameter $q=\delta\Phi(t)/\Phi_0$. Here $\delta \Phi\equiv \Phi - \Phi_0/2$ and $\Phi_0 =h/2e$ is the superconducting flux quantum.
\begin{figure}
\centering
\includegraphics [width=0.5\columnwidth] {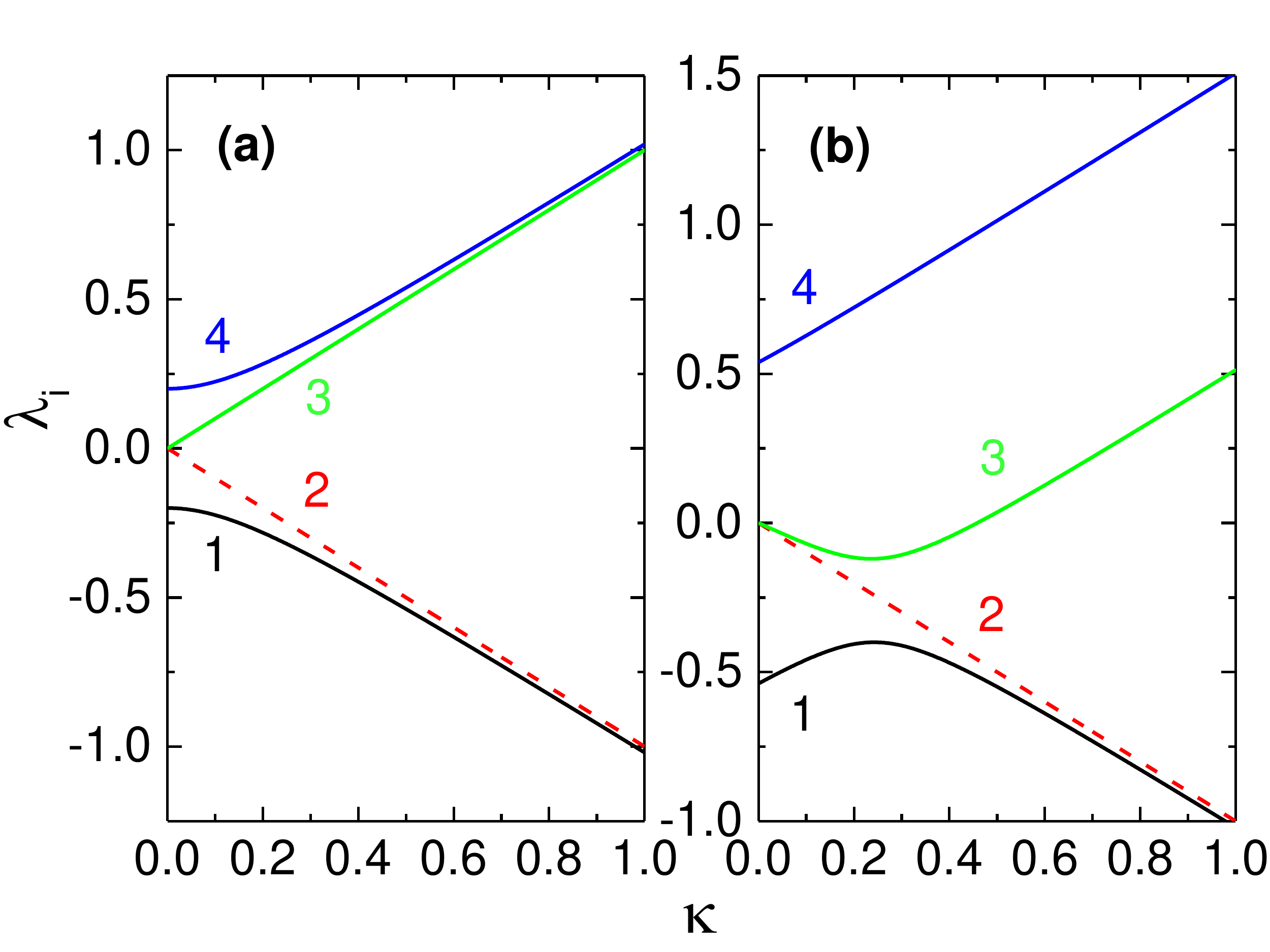}
\caption{Dependence of eigenenergies $\lambda_i$ $(i=1,2,3,4)$ on the coupling parameter $\kappa$ for different values of dimensionless flux $q$, (a) $q=0$, (b) $q=0.25$. In both panels $\Delta=0.1$.}
\label{fig2}
\end{figure}
For the noise, we assume linear coupling as $H_{{c}j,{{\rm B}}} = g_{\rm B} i_{n,{\rm B}}(t)\sigma_{z,j}$, and similarly for other couplings. Here, $g_{{\rm B}}=\frac{E_0 M_{\rm B}}{\Phi_0}$ is the coupling of the reservoir ${\rm B}$ to each qubit, and $i_{n,{\rm B}}(t)$ is the noise current of the cold/hot environment. %\cred{$M_{C/H}=M_{1C/1H}+M_{2C/2H}$}
The interaction Hamiltonian $H_{12}$ between two qubits which is the energy corresponding to the standard inductive coupling of two current loops, is given by 
\begin{equation} \label{h3}
H_{12}=M I_1 I_2,
\end{equation}
where $I_j=\frac{\partial H_{{\rm Q}j}}{\partial \Phi_j} = - \frac{E_j}{\Phi_0} \sigma_{z,j}$ are the current operators of the two qubits. Therefore, with $\gamma = M E_1E_2/\Phi_0^2$, we have
\begin{equation} \label{h5}
H_{12} = \gamma \sigma_{z,1} \sigma_{z,2}
\end{equation}
and the total Hamiltonian of the two coupled qubits, $H_S = H_{{\rm Q}1}+H_{{\rm Q}2}+H_{12}$, reads
\begin{equation} \label{h6}
H_S =  -\sum_{j=1,2} E_j (\Delta_j \sigma_{x,j} + q\sigma_{z,j})+\gamma \sigma_{z,1} \sigma_{z,2}.
\end{equation}
For quantitative analysis, we use the Bell basis of maximally entangled states $\{\frac{1}{\sqrt{2}}(|0_10_2\rangle +|1_11_2\rangle),~\frac{1}{\sqrt{2}}(|0_10_2\rangle -|1_11_2\rangle),~\frac{1}{\sqrt{2}}(|0_11_2\rangle +|1_10_2\rangle),~\frac{1}{\sqrt{2}}(|0_11_2\rangle -|1_10_2\rangle)\}$. Considering the fully symmetric system, $E_0\equiv E_1=E_2$, and $\Delta \equiv \Delta_1=\Delta_2$, and the relevant matrices in this basis (details are provided in the Appendix) the Hamiltonian is given by
\begin{eqnarray} \label{hamilton1}
H_{\rm S} = E_0\left(
\begin{array}{cccc}
\kappa & -2q &-2\Delta & 0 \\ -2q & \kappa & 0 & 0 \\ -2\Delta & 0 & -\kappa & 0 \\ 0 & 0 & 0& -\kappa
\end{array}
\right),
\end{eqnarray}
where $\kappa = \gamma /E_0$.
The eigenenergies $\lambda_i$ (normalized by $E_0$) of the Hamiltonian \eqref{hamilton1} are given by
\begin{equation} \label{ee1}
(\kappa +\lambda_i)[\lambda_i^3-\kappa\lambda_i^2-(\kappa^2+4q^2+4\Delta^2)\lambda_i+\kappa(\kappa^2-4q^2+4\Delta^2)]=0.
\end{equation}
Out of the eigenenergies, $\lambda_2=-\kappa$ represents a "protected state" as will become evident in what follows, and the rest of the energies $\lambda_1,\lambda_3,\lambda_4$ form a three-state system. Figure \ref{fig2}a,b displays these eigenenergies as a function of $\kappa$ for two different values of $q$. It is vivid that coupling influences all energy levels and splits the initially degenerate levels $(2$ and $3)$. Introducing flux to the system (varying $q$), modifies the three level structure.
It is then straightforward to find the corresponding normalized eigenstates by solving a set of three linear equations, yielding
\begin{eqnarray} \label{es1}~
|i\rangle = \frac{\big{[}(\kappa^2-\lambda_i^2),2q(\kappa+\lambda_i), -2\Delta(\kappa-\lambda_i),0\big{]}^{\rm T}}{\sqrt{(\kappa^2-\lambda_i^2)^2+4q^2(\kappa+\lambda_i)^2+4\Delta^2(\kappa-\lambda_i)^2}},
\end{eqnarray}
for $i=1,3,4,$ and $|2\rangle=[0~0~0~1]^{\rm T}.$
The transition rates between eigenstates $i$ and $j$ having an energy spacing $E_{ij}$ of the coupled qubits are sensitive to current noise produced by two baths at the frequency $\omega_{ij}=E_{ij}/\hbar$ and can be calculated from Fermi's golden rule. These rates are given by
\begin{equation}
\Gamma_{i \rightarrow j, {\rm B}}=\frac{g^2}{\hbar^2} |\langle i|\sigma_z^{\Sigma}|j\rangle|^2 S_{I,{\rm B}}(\omega_{ij}),
\end{equation}
where $g_{\rm C}=g_{\rm H}=g$, $\sigma_z^{\Sigma}=\sigma_{z,1}+\sigma_{z,2}$, $\omega_{ij}=E_{ij}=E_0(\lambda_i -\lambda_j)/\hbar$ and $S_{I,{\rm B}}(\omega_{ij})=[1+Q_{\rm B}^2(\frac{\omega_{ij}}{\omega_{LC,{\rm B}}}-\frac{\omega_{LC,{\rm B}}}{\omega_{ij}})^2]^{-1} \frac{2\hbar\omega_{ij}}{R_{\rm B}(1-e^{-\hbar\omega_{ij}/k_BT_{\rm B}})}$ is the unsymmetrized noise spectrum. Here,  $\omega_{LC,{\rm B}}= 1/\sqrt{L_{\rm B}C_{\rm B}}$ and $Q_{\rm B}=\sqrt{L_{\rm B}/C_{\rm B}}/R_{\rm B}$ are the bare resonance angular frequency and the quality factor of each $LC$-circuit. Now we indeed see that $|2\rangle$ is a protected in this set-up state since $\sigma_z^\Sigma|2\rangle\equiv0$, and hence $\Gamma_{2\rightarrow i,\rm B}=\Gamma_{i\rightarrow 2,\rm B}=0$ for all $i$ and $\rm B$.
\begin{figure}
\centering
\includegraphics [width=\columnwidth] {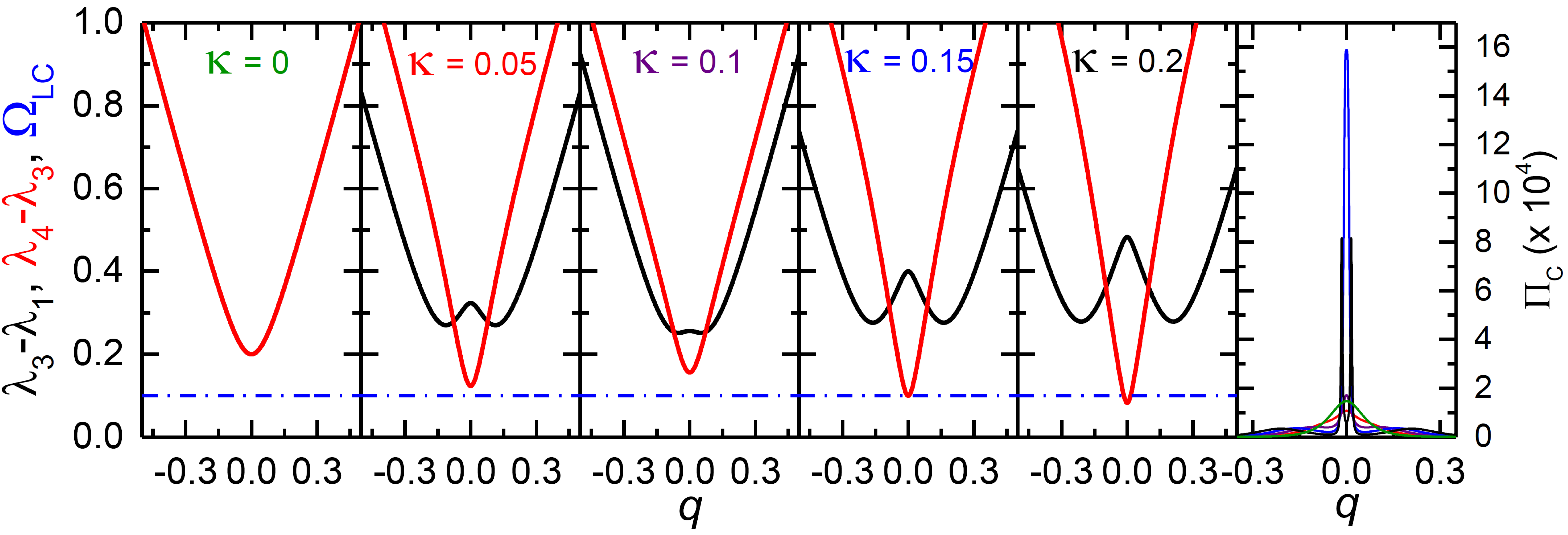}
\caption{(left) The variation of energy level separations $\lambda_3-\lambda_1$ (black lines) and $\lambda_4-\lambda_3$ (red lines) as a function of drive for five different values of coupling. Dot-dashed line corresponds to dimensionless resonance frequency $\Omega_{LC}=0.1$ of the $LC$ resonators. (right) Flux $q$ dependence of the transmitted dimensionless power $\Pi_{\rm C}=\hbar P_{\rm C}/E_0^2$ for different values of coupling. The colors of the labels from left panels correspond to the related ones in the right panel. The parameters $E_0/k_B T_{{\rm C}}=20$, $E_0/k_B T_{{\rm H}}=5$, $Q=Q_{\rm C}=Q_{\rm H}=10$, $g=g_{\rm C}=g_{\rm H}=1$, and $\Delta=0.1$ were used for all panels.}
\label{fig3}
\end{figure}
In steady state, the master equation (see Appendix) yields the populations $\rho_{ii}$ of the levels as 
\begin{widetext}
\begin{eqnarray} \label{diagrho}
&&\rho_{ii}=\frac{\Gamma_{j\rightarrow k}\Gamma_{k\rightarrow i}+\Gamma_{j\rightarrow i}\Gamma_{k\rightarrow i}+\Gamma_{j\rightarrow i}\Gamma_{k\rightarrow j}}{(\Gamma_{3\rightarrow 1}+\Gamma_{3\rightarrow 4})(\Gamma_{1\rightarrow 4}+\Gamma_{4\rightarrow 1})+(\Gamma_{1\rightarrow 4}+\Gamma_{3\rightarrow 1})\Gamma_{4\rightarrow 3}+\Gamma_{1\rightarrow 3}(\Gamma_{3\rightarrow 4}+\Gamma_{4\rightarrow 1}+\Gamma_{4\rightarrow 3})}\nonumber \\
%&&\rho_{33}=\frac{\Gamma_{1\rightarrow 4}\Gamma_{4\rightarrow 3}+\Gamma_{1\rightarrow 3}(\Gamma_{4\rightarrow 1}+\Gamma_{4\rightarrow 3})}{(\Gamma_{3\rightarrow 1}+\Gamma_{3\rightarrow 4})(\Gamma_{1\rightarrow 4}+\Gamma_{4\rightarrow 1})+(\Gamma_{1\rightarrow 4}+\Gamma_{3\rightarrow 1})\Gamma_{4\rightarrow 3}+\Gamma_{1\rightarrow 3}(\Gamma_{3\rightarrow 4}+\Gamma_{4\rightarrow 1}+\Gamma_{4\rightarrow 3})}\nonumber \\
%&&\rho_{44}=\frac{\Gamma_{1\rightarrow 3}\Gamma_{3\rightarrow 4}+\Gamma_{1\rightarrow 4}(\Gamma_{3\rightarrow 1}+\Gamma_{3\rightarrow 4})}{(\Gamma_{3\rightarrow 1}+\Gamma_{3\rightarrow 4})(\Gamma_{1\rightarrow 4}+\Gamma_{4\rightarrow 1})+(\Gamma_{1\rightarrow 4}+\Gamma_{3\rightarrow 1})\Gamma_{4\rightarrow 3}+\Gamma_{1\rightarrow 3}(\Gamma_{3\rightarrow 4}+\Gamma_{4\rightarrow 1}+\Gamma_{4\rightarrow 3})},
\end{eqnarray}
\end{widetext}
where $\Gamma_{i\rightarrow j}$ is the total transition rate $(\Gamma_{i\rightarrow j}=\Gamma_{i\rightarrow j,{\rm C}}+\Gamma_{i\rightarrow j,{\rm H}})$ due to both the baths from eigenstate $i$ to $j$. The indices assume values $(i,j,k)=(1,3,4)$, and their cyclic permutations.
Equation \eqref{diagrho} applies, when the system is initialized in the subspace $\{|1\rangle,|3\rangle,|4\rangle\}$.
The expression of power to bath ${\rm B}$ can be written in the form
\begin{equation} \label{power1}
P_{\rm B} = \frac{2i}{\hbar^2} g^2E_0\Delta \sum_{k,l} \rho_{kk}\sigma_{y,kl}^\Sigma \sigma_{z,lk}^\Sigma S_{I,{\rm B}}(\omega_{kl}).
\end{equation} 
It is easy to show that $\frac{\sigma_{y,kl}^\Sigma}{\sigma_{z,kl}^\Sigma}=\frac{\lambda_k-\lambda_l}{2\Delta}i$, where $\sigma_{y,kl}^\Sigma=\langle k|\sigma_{y,1}+\sigma_{y,2}|l\rangle$ and $\sigma_{z,kl}^\Sigma=\langle k|\sigma_{z,1}+\sigma_{z,2}|l\rangle$. So, we can simplify the expression of power in a two-qubit system to
\begin{equation} \label{final.power}
P_{\rm B} = \sum_{k,l} \rho_{kk} E_{kl}\Gamma_{k\rightarrow l,{\rm B}}. 
\end{equation}
\begin{figure}[h]
\centering
\includegraphics [width=\columnwidth] {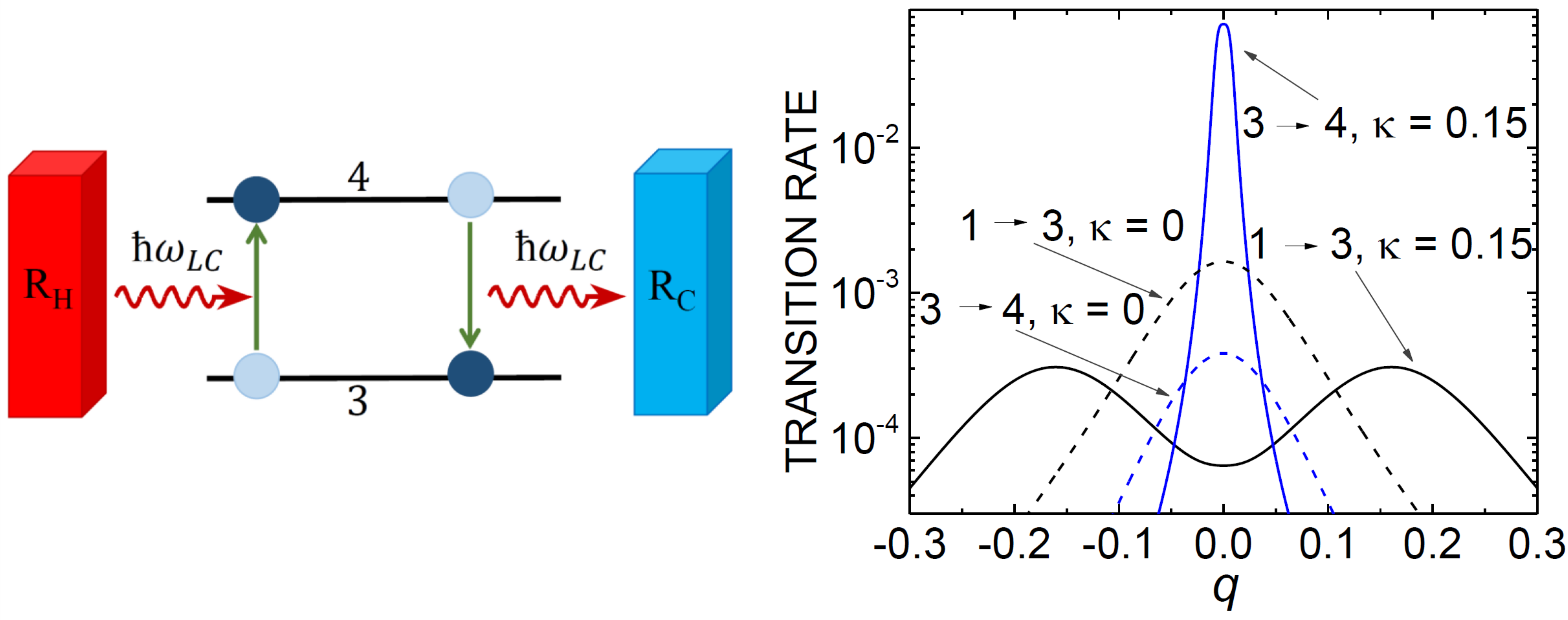}
\caption{(left) Scheme of the energy transport mechanism when resonator frequency matches the energy level separation between the two excited states 3 and 4. (right) Total transition rates ($\rho_{ii}\Gamma_{i \rightarrow j}$) between three energy levels 1, 3, and 4 as a function of $q$. The dashed lines illustrate transition rates in the case of decoupled qubits $(\kappa=0)$ while the solid lines are for coupled qubits $(\kappa=0.15)$ corresponding to the maximum power transfer. The parameters are $E_0/k_B T_{\rm H}=5$, $E_0/k_B T_{\rm C}=20$, $Q=Q_{\rm C}=Q_{\rm H}=10$, $g=g_{\rm C}=g_{\rm H}=1$, $\Delta=0.1$.
\label{fig4}}
\end{figure}
\section{Role of energy level separation in Transmitted power for entangled qubits}
In left panels of Fig. \ref{fig3}a, we plot the energy separation, $\lambda_3-\lambda_1$ (black lines) and $\lambda_4-\lambda_3$ (red lines), for five different values of coupling $\kappa$ as a function of flux bias $q$. Dot-dashed line represents the dimensionless resonance frequency $\Omega_{LC}=\frac{\hbar\omega_{LC}}{E_0}$. The dependence of transmitted power on flux $q$ for these values of coupling is shown in the right panel of Fig. \ref{fig3}. With no coupling ($\kappa=0$) the two level separations are equal: $\lambda_3-\lambda_1=\lambda_4-\lambda_3=2\sqrt{q^2+\Delta^2}$. By increasing coupling $\kappa$ from $0$ to $0.1$, the power initially decreases because of increasing difference between energy levels $1$ and $3$. Further increasing $\kappa$ the power increases abruptly obtaining the maximum at $\kappa=0.15$. This is because, here the $\lambda_4-\lambda_3$ separation meets the resonance frequency at $q=0$. By further increasing coupling, $\lambda_4-\lambda_3$ will cross $\Omega_{LC}$ at two values of $q$ and leads to two peaks in power, but the maxima decrease because of larger $\lambda_3-\lambda_1$. It is clear that the $q$ dependence of $\lambda_4-\lambda_3$  becomes stronger when $\kappa$ increases from 0 to 0.2. This is the feature that determines the bandwidth (in $q$) of the filter. 

\begin{figure}
\centering
\includegraphics [width=\columnwidth] {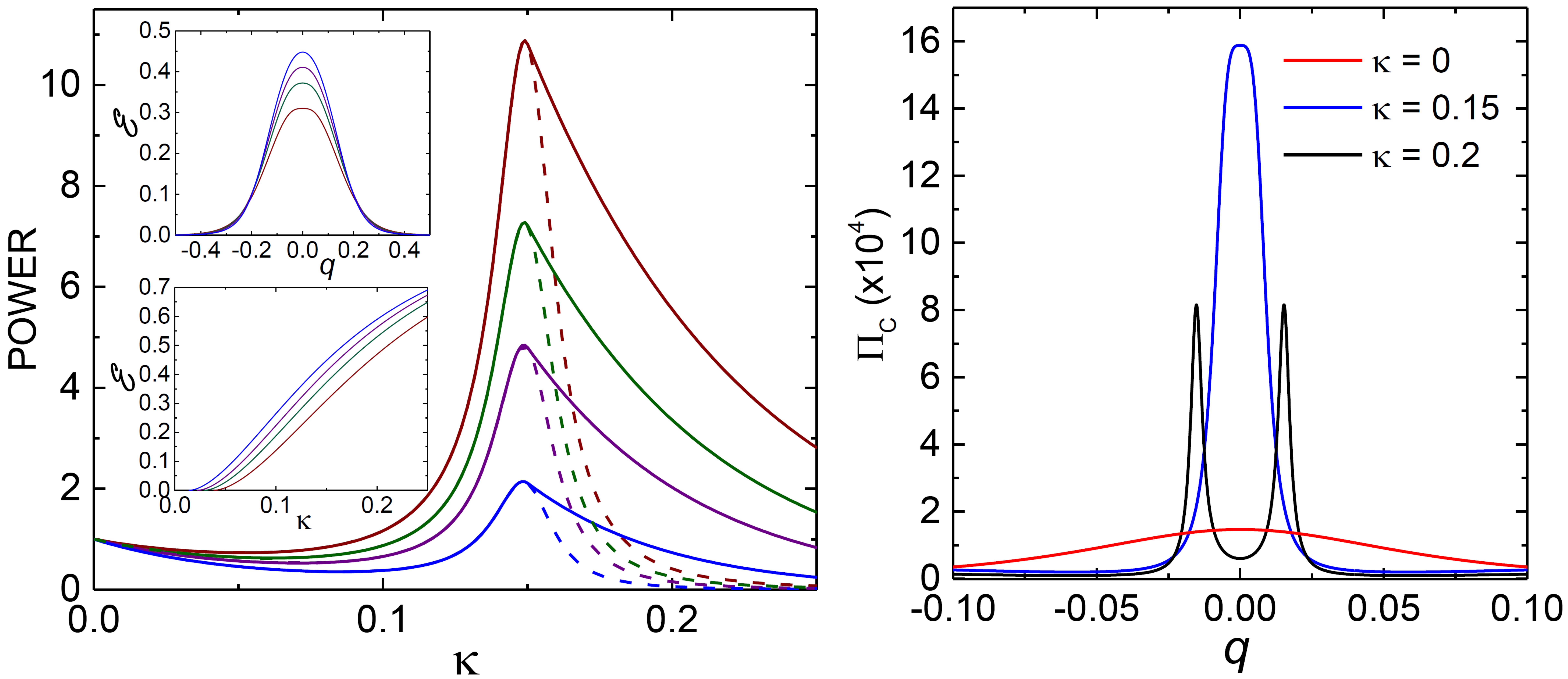}
\caption{(left) The variation of transmitted power normalized by that at $\kappa=0$ as a function of coupling between the qubits for different bath temperatures. The solid lines display maximum power with respect to $q$ for each value of coupling while the dashed lines depict the power at $q=0$. (right) Dependence of the transmitted dimensionless power $\Pi_{\rm C}$ on flux $q$ for three different values of coupling: $\kappa=0$ (red line), $\kappa=0.15$ (blue line), and $\kappa=0.2$ (black line). Insets: Entanglement as functions of $q$ (top) and coupling (bottom) for the same temperatures as in the main frame. The parameters for all the panels are $E_0/k_B T_{\rm C}=20$, $Q=Q_{\rm C}=Q_{\rm H}=10$, $g=g_{\rm C}=g_{\rm H}=1$, $\Delta=0.1$, $E_0/k_B T_{\rm H}=5$ (brown lines), $E_0/k_B T_{\rm H}=6.25$ (green lines), $E_0/k_B T_{\rm H}=7.5$ (purple lines), and $E_0/k_B T_{\rm H}=10$ (blue lines).
\label{fig5}}
\end{figure}

The mechanism of increased power with $\kappa$ is shown schematically in Fig. \ref{fig4} (left): the total transition rates (relaxation and excitation) increase dramatically as shown in Fig. \ref{fig4} (right). There the relevant transition rates between levels for decoupled and coupled $(\kappa=0.15)$ qubits are depicted. Numerical results of the power $P_{\rm C}$ as a function of coupling at different bath temperatures are displayed in the left panel of Fig. \ref{fig5}. Here we have fixed the temperature of the cold bath and vary that of the hot one. Based on this figure we see the obvious result: the higher the difference between the temperatures, the higher the power transferred to the cold bath. The precise evolution of power as a function of coupling $\kappa$ is a result of the interplay of the level structure of the coupled qubits and the parameters of the baths. The quality factors do not influence the maximum of the transmitted power because the transition rates do not depend on $Q_{\rm B}$ at resonance. Yet in our example, high $Q_{\rm B}$ decreases $P_{\rm C}$ at $\kappa=0$ strongly. This is because under these conditions, the $LC$-circuit is not in resonance with the qubits at any value of $q$. 

Given the prominent role of the coherent coupling in determining the boost in the emitted power it is relevant to explore if/how quantum correlations are the underlying reason for this effect. As we will see in the following, entanglement is indeed present and plays some role although it is not in a direct correspondence with the enhancement factor. Entanglement plays an important role in quantum information processing \cite{saro,horodecki}. Bipartite systems formed of two coupled qubits have attracted lots of attention in this respect \cite{saro,steffen, narla}. In order to assess the degree of entanglement of the coupled qubits in our set-up, we adopt the concurrence introduced by Wootters \cite{wootters98} as $C(\rho)={\rm max}\{0,\sqrt{\mu_1}-\sqrt{\mu_2}-\sqrt{\mu_3}-\sqrt{\mu_4}\}$, where $\mu_i~(i=1,2,3,4)$ are the eigenvalues  arranged in decreasing order of the non-Hermitian matrix $R=\rho \tilde \rho$. Here $\tilde \rho = (\sigma_{y,1}\otimes\sigma_{y,2})\rho^* (\sigma_{y,1}\otimes\sigma_{y,2})$ and $\rho^*$ denotes the complex conjugation of $\rho$ and $\sigma_{y,1}$ and $\sigma_{y,2}$ are the relevant Pauli matrices.
Concurrence varies from $0$ for a completely unentangled state to $1$ for a maximally entangled one. The entanglement $\mathcal E(\rho)$ is given by $\mathcal E(\rho)=-x\log_2(x)-(1-x)\log_2(1-x)$, where $x=[1+\sqrt{1-C^2}]/2$. The dependence of entanglement on flux and coupling is shown in the inset of the left panel of Fig. \ref{fig5} (top and bottom), respectively, for different bath temperatures. It is clear that at optimal value of coupling $(\kappa=0.15)$, the qubits are highly entangeled up to 30\%. 
Finally it is fair to note that the enhancement of power when coupling the qubits is somewhat case-dependent. For the presented case of $\Omega_{LC} < 2\Delta$ this effect is prominent, since with increasing coupling one eventually meets the resonance condition discussed. The numerics shows that in the opposite regime, $\Omega_{LC} > 2\Delta$, there is also weak enhancement, due to matching of $\Omega_{LC} = \lambda_3-\lambda_1$ condition at finite $\kappa$. Yet, when the system is at the resonance $\Omega_{LC} = \lambda_3-\lambda_1 = \lambda_4-\lambda_3$ already at $\kappa=0$, the transmitted maximum power decreases monotonically as a function of $\kappa$.

\section{Better switching with coupled qubits?}
In order to assess the bandwidth of the filter $\delta q$, we calculate the curvature of the eigenenergies with respect to $q$, i.e. $w\equiv \big{[}\frac{d^2(\lambda_4-\lambda_3)}{dq^2} \big{]}^{-1}|_{q=0}$. Differentiating the expression of eigenenergies of the Hamiltonian Eq. (\ref{ee1}) twice with respect to $q$ and putting $\frac{d\lambda_i}{dq}=0$ at $q=0$, we have
\begin{equation}
\frac{d^2\lambda_i}{dq^2}=\frac{8(\kappa+\lambda_i)}{3\lambda_i^2-2\kappa\lambda_i-\kappa^2-4\Delta^2}.
\end{equation}
This yields for the width
\begin{equation} \label{filter}
%w(\kappa)\equiv \big{[}\frac{d^2(\lambda_4-\lambda_3)}{dq^2} \big{]}^{-1}|_{q=0}=\frac{1}{4}\big{[}(\frac{\kappa+\sqrt{\kappa^2+4\Delta^2}}{(\sqrt{\kappa^2+4\Delta^2}-\kappa)\sqrt{\kappa^2+4\Delta^2}}+\frac{\kappa}{\Delta^2})\big{]}^{-1}.
w(\kappa)=\frac{1}{4}\big{[}(\frac{\kappa+\sqrt{\kappa^2+4\Delta^2}}{(\sqrt{\kappa^2+4\Delta^2}-\kappa)\sqrt{\kappa^2+4\Delta^2}}+\frac{\kappa}{\Delta^2})\big{]}^{-1}.
\end{equation}
For $\kappa=0$, we have $w(0)=\big{[}\frac{d^2(\lambda_4-\lambda_3)}{dq^2} \big{]}^{-1}|_{q=0}=\frac{\Delta}{2}$. We want to compare this value to that at optimal coupling where $\lambda_4-\lambda_3 =\Omega_{LC}$ at $q=0$, which occurs according to Eqs. \eqref{enk0} at $\sqrt{\kappa^2+4\Delta^2}-\kappa =\Omega_{LC}$, i.e. at
\begin{equation} \label{kappaopt}
\kappa_{\rm opt}= \frac{4\Delta^2-\Omega_{LC}^2}{2\Omega_{LC}},
\end{equation}
where $\Omega_{LC}=\hbar\omega_{LC}/E_0$ is the dimensionless resonator frequency. In Fig. \ref{fig3}, $\Delta = \Omega_{LC}$, which yields $\kappa_{\rm opt}= 3/2$ for the optimum where the level separation $\lambda_4-\lambda_3$ meets the resonance frequency at $q=0$; this is consistent with our numerics. Inserting Eq. \eqref{kappaopt} into Eq. \eqref{filter}, we have
\begin{equation} \label{filter1}
w(\kappa_{\rm opt})=\frac{\Omega_{LC}}{4}\frac{2\Delta^2(4\Delta^2+\Omega_{LC}^2)}{32\Delta^4-\Omega_{LC}^4}.
\end{equation}
The expression above allows us to see how much one can improve the bandwidth of filtering by coupling the two qubits. If we divide expression \eqref{filter1} by the corresponding width for decoupled qubits in the example of Fig. 3, we obtain an improvement in the filter bandwidth as 
\begin{equation} \label{filter2}
w(\kappa_{\rm opt})/(\Omega_{LC}/2) = 5/31,
\end{equation}
i.e. it is narrower by factor 6.2. To make our argument more general, we compare the bandwidth at maximum power of decoupled and optimally coupled qubits. To start with, we assume a fixed value for $\Omega_{LC}$. For decoupled qubits the maximum power is obtained at the resonance condition $\Delta_{0}= \Omega_{LC}/2$. In this case, we have for the decoupled qubits $\big{[}\frac{d^2(\lambda_4-\lambda_3)}{dq^2} \big{]}^{-1}|_{q=0} = \Omega_{LC}/4$. The ratio $r$ of the bandwidth at maximum power for optimally coupled and fully decoupled qubits, is given by 
\begin{equation} \label{filterf}
r=\frac{2\Delta^2(4\Delta^2+\Omega_{LC}^2)}{32\Delta^4-\Omega_{LC}^4}.
\end{equation} 
It is clear that one benefits of setting $\Delta \gg \Omega_{LC}$, and for $\Delta/\Omega_{LC}\rightarrow \infty$ we obtain the ultimate improvement of $r=1/4$. (In our numerical example in Fig. \ref{fig3} this improvement would be $10/31$). 

We separately show the transmitted power for decoupled $(\kappa=0)$ and coupled $(\kappa=0.15$ and $\kappa=0.2)$ qubits as a function of flux bias $q$ in a smaller range in the right panel of Fig. \ref{fig5}. Here based on numerical results for transmitted power, the ratio of the full width at half maximum (FWHM) of the peaks at coupled $(\kappa=0.15)$ and decoupled $(\kappa=0)$ cases equals 6.8, which is consistent with our analytical result 6.2 given above for the bandwidth of the filter. Broader and weaker peak of single qubit in comparison with the sharper peak of coupled qubits indicates that coupled qubits provide a more selective heat switch.

\section{Conclusion}

We have introduced a design of a quantum heat switch based on two coupled qubits. We present explicit results of the amount of heat transferred between the hot and the cold baths. We have shown that at optimal value of coupling, when the energy level difference of the two excited states crosses the resonance frequency of the environment, the corresponding transition rates, and thus the transmitted power increase dramatically, in our example by an order of magnitude. As a result, we achieve a four times narrower bandwidth at maximum power of the heat switch based on coupled qubits as compared to that of two independent ones.

\section*{Appendix}
\setcounter{section}{0}
\setcounter{equation}{0}
\setcounter{figure}{0}
\setcounter{table}{0}
\renewcommand{\theequation}{A\arabic{equation}}
\renewcommand{\thefigure}{A\arabic{figure}}

\section{Coupling to the baths}
In general, we write the master equation for the coupled qubit system in the absence of active time-dependent driving as
\begin{equation} \label{diss1}
\dot \rho = \mathcal L(\rho),
\end{equation}
where the rhs arises from coupling to the baths and can be written in the rotating wave approximation and for the case $g_{C_1}=g_{C_2}\equiv g_C$, and $g_{H_1}=g_{H_2}=g_H$ as
\begin{eqnarray}
\mathcal L(\rho)_{kl} =-\sum_{B=C,H} \frac{g_B^2}{2\hbar^2}\bigg{[}\rho_{kl}\sum_{j=1}^4 \big\{|\langle j| \sigma^\Sigma_z|l\rangle|^2 S_B(\omega_{lj})+
|\langle j| \sigma^\Sigma_z|k\rangle|^2 S_B(\omega_{kj})\big\}-2\delta_{kl} \sum_{j=1}^4 \rho_{jj}|\langle j|\sigma^\Sigma_z|k\rangle|^2 S_B(\omega_{jk})\bigg{]}.
\end{eqnarray} 
Here $\sigma^\Sigma_z \equiv \sigma_{z,1}+\sigma_{z,2}$, and we find the rates $\Gamma_{k\rightarrow j, B} = \frac{g_B^2}{\hbar^2} |\langle j| \sigma^\Sigma_z|k\rangle|^2 S_B(\omega_{kj})$, yielding
\begin{eqnarray} \label{finalL}
\mathcal L(\rho)_{kl} =\delta_{kl} \sum_{j=1}^4 \rho_{jj}\Gamma_{j\rightarrow k}-\frac{1}{2}\rho_{kl}\sum_{j=1}^4 (\Gamma_{l\rightarrow j}+\Gamma_{k\rightarrow j}).
\end{eqnarray}
$\Gamma_{k\rightarrow j}=\Gamma_{k\rightarrow j,C}+\Gamma_{k\rightarrow j,H}$ is the total transition rate from state $k$ to state $j$.
For the diagonal elements we then have the master equation in the form
\begin{eqnarray} \label{finalme}
\dot\rho_{kk} = \sum_{j=1}^4 ( \rho_{jj}\Gamma_{j\rightarrow k}-\rho_{kk} \Gamma_{k\rightarrow j}).
\end{eqnarray}

\section{Power}
The power to the cold reservoir, $P_C$, can be obtained in the standard way as $P_C =-\langle \dot H_{S,\rm C} \rangle$, where $\dot H_{S,\rm C} =\frac{i}{\hbar}[H_{\rm c,C},H_S]$ is the operator of power to the coupled qubits from the cold bath, and $H_{\rm c,C}=H_{\rm c1,C} + H_{\rm c2,C}=g(\sigma_{z,1}+\sigma_{z,2})i_{n,C}(t)$. We then find that
\begin{equation} \label{power1}
\dot H_S = \frac{2g}{\hbar}E_0\Delta (\sigma_{y,1}+\sigma_{y,2})i_{nC}(t)=\frac{2g}{\hbar}E_0\Delta \sigma_y^\Sigma i_{nC}(t).
\end{equation}
We write with standard notations the power transmitted between the hot and cold baths in linear response as
\begin{equation} \label{power2}
P_C=\frac{i}{\hbar} \int_{-\infty}^t dt' \langle [\dot H_{S,\rm C,I}(t),H_{{\rm cC},I}(t')]\rangle,
\end{equation}
where the subscript $I$ refers to the interaction picture. After a straightforward calculation we obtain
\begin{equation} \label{power3}
P_C = \frac{2i}{\hbar^2} g^2E_0\Delta \sum_{k,l} \rho_{kk}\sigma_{y,kl}^\Sigma\sigma_{z,lk}^\Sigma S_{I,C}(\omega_{kl}).
\end{equation} 
By the identity given in the main text we can finally show that this expression is equivalent to
\begin{equation} \label{final.power}
P_C = \sum_{k,l} \rho_{kk} E_{kl}\Gamma_{k\rightarrow l,C}. 
\end{equation}

\section{matrices, eigenenergies, and eigenstates}
The relevant matrices in Bell basis are given by
\begin{eqnarray} \label{sx1}
\sigma_{x,1} = \left(
\begin{array}{cccc}
0&0&1&0 \\0&0&0&-1 \\ 1&0&0&0 \\ 0&-1&0&0
\end{array}
\right),
\sigma_{x,2} = \left(
\begin{array}{cccc}
0&0&1&0 \\0&0&0&1 \\ 1&0&0&0 \\ 0&1&0&0
\end{array}
\right),\nonumber
\end{eqnarray}
\begin{eqnarray}
\sigma_{y,1} = \left(
\begin{array}{cccc}
0&0&0&i \\0&0&-i&0 \\ 0&i&0&0 \\ -i&0&0&0
\end{array}
\right),
\sigma_{y,2} = \left(
\begin{array}{cccc}
0&0&0&-i \\0&0&-i&0 \\ 0&i&0&0 \\ i&0&0&0
\end{array}
\right),\nonumber
\end{eqnarray}
\begin{eqnarray}
\sigma_{z,1} = \left(
\begin{array}{cccc}
0&1&0&0 \\1&0&0&0 \\ 0&0&0&1 \\ 0&0&1&0
\end{array}
\right),
\sigma_{z,2} = \left(
\begin{array}{cccc}
0&1&0&0 \\1&0&0&0 \\ 0&0&0&-1 \\ 0&0&-1&0
\end{array}
\right).
\end{eqnarray}
\subsection{Decoupled qubits}
When $\kappa=0$, we find the eigenenergies in units of $E_0$ and eigenstates as
\begin{eqnarray} \label{ens0}
\lambda_1=-2\sqrt{q^2+\Delta^2},
\lambda_2=0,
\lambda_3=0,
\lambda_4=+2\sqrt{q^2+\Delta^2}.
\end{eqnarray}
The corresponding eigenstates are
\begin{eqnarray} \label{es01}
|1\rangle = \frac{1}{\sqrt{2}}\left(
\begin{array}{cccc}
1 \\ q/\sqrt{q^2+\Delta^2} \\ \Delta /\sqrt{q^2+\Delta^2}\\ 0
\end{array}
\right),
|2\rangle = \left(
\begin{array}{cccc}
0 \\ 0 \\0\\ 1
\end{array}
\right),
|3\rangle = \left(
\begin{array}{cccc}
0 \\ \Delta/\sqrt{q^2+\Delta^2} \\ -q /\sqrt{q^2+\Delta^2}\\ 0
\end{array}
\right),
|4\rangle = \frac{1}{\sqrt{2}}\left(
\begin{array}{cccc}
1 \\ -q/\sqrt{q^2+\Delta^2} \\ -\Delta /\sqrt{q^2+\Delta^2}\\ 0
\end{array}
\right).
\end{eqnarray}

\subsection{Coupled qubits, $q=0$}
Another regime, where we find explicit results for eigenenergies and eigenstates is that with finite $\kappa$ but at $q=0$. We have then the solutions
\begin{eqnarray} \label{enk0}
\lambda_1=-\sqrt{\kappa^2+4\Delta^2},
\lambda_2=-\kappa,
\lambda_3=+\kappa,
\lambda_4=+\sqrt{\kappa^2+4\Delta^2}.
\end{eqnarray}
The corresponding eigenstates are
\begin{eqnarray} \label{ek01}
|1\rangle = \frac{1}{\sqrt{1+(\frac{\kappa+\sqrt{\kappa^2+4\Delta^2}}{2\Delta})^2}}\left(
\begin{array}{cccc}
1 \\ 0 \\ \frac{\kappa+\sqrt{\kappa^2+4\Delta^2}}{2\Delta}\\ 0
\end{array}
\right), 
|2\rangle = \left(
\begin{array}{cccc}
0 \\ 0 \\ 0\\ 1
\end{array}
\right),
|3\rangle = \left(
\begin{array}{cccc}
0 \\ 1 \\ 0\\ 0
\end{array}
\right),
|4\rangle = \frac{1}{\sqrt{1+(\frac{\kappa+\sqrt{\kappa^2+4\Delta^2}}{2\Delta})^2}}\left(
\begin{array}{cccc}
\frac{\kappa+\sqrt{\kappa^2+4\Delta^2}}{2\Delta} \\ 0 \\ -1\\ 0
\end{array}
\right).\nonumber\\
\end{eqnarray} 

We acknowledge Yuri Galperin and Dmitry Golubev for useful discussions. The work was supported by the Academy of Finland (grants 272218 and 284594).

\end{document}